# Design and numerical performance analysis of efficient $Ag_3TaX_4$ (X = S, Se, Te) thin film solar cells

Md. Nahid Hasan[1], Tanvir Ahmed[2], Md. Abdur Rashid[3], Tanzina Rahman[1], Dinesh Pathak[4] and Jaker Hossain[1*]

[1]*Photonics & Advanced Materials Laboratory, Department of Electrical and Electronic Engineering, University of Rajshahi, Rajshahi 6205, Bangladesh.*

[2]*Department of Electrical & Electronic Engineering, First Capital University of Bangladesh, Chuadanga 7200, Bangladesh.*

[3]*Department of Physics, University of Rajshahi, Rajshahi 6205, Bangladesh.*

[4]*Department of Physics, The University of the West Indies, St. Augustine Campus, Trinidad and Tobago.*

**Abstract**

Silver-based ternary chalcogenides have recently emerged as promising absorber materials for thin film photovoltaics. Nevertheless, their photovoltaic performance in complete device architectures has not yet been systematically explored. In this work, three-dimensional (3D) n-CdS/p-$Ag_3TaX_4$ (X = S, Se, Te)/$p^+$-GeS thin-film solar cells have been designed and numerically investigated using the Semiconductor Module of COMSOL Multiphysics. Herein, the various performance matrices of the proposed devices have been analysed in accordance with the changing of depth, carrier, and defect concentration in each layer of the structures. The optimized $Ag_3TaS_4$-based device delivers a power conversion efficiency, PCE of 24.66%, open circuit voltage, $V_{OC}$ of 1.4V, short circuit current density, $J_{SC}$ of 20.68 mA/cm$^2$, and fill factor, FF of 85.16%. The $Ag_3TaSe_4$-based solar cell exhibits the PCE of 28.1% with $V_{OC}$ = 1.19V, $J_{SC}$ = 27.0 mA/cm$^2$, and FF = 87.44%. The $Ag_3TaTe_4$ solar device shows a PCE of 27.56% with a $V_{OC}$ of 0.88 V, $J_{SC}$ of 36.14 mA/cm$^2$, fill factor of 86.65%. These results provide a deeper insight into device operation and offer practical design guidelines for fabricating efficient $Ag_3TaX_4$ (X = S, Se, T e)-based novel next-generation solar cells.

**Keywords:** $Ag_3TaX_4$ (X=S, Se, Te), CdS, GeS, Solar cell, efficient, COMSOL.

## 1. Introduction

The rapid growth in global energy consumption, together with growing concerns over climate and greenhouse gas emissions, has intensified the search for sustainable, environmentally benign energy technologies. Solar energy is a clean and sustainable resource that produces no harmful emissions, making it one of the most fascinating, viable, and inexpensive energy sources. The amount of sunlight that reaches the Earth in an hour is proportional to the energy used by humans in a year [1-2]. Hence, the conversion of photo energy into electrical power remains one of the most dependable and scalable strategies for meeting future global energy demands. Among all renewable resources, solar cells (SCs) offer the greatest potential, and they can convert light into functioning electricity through photovoltaic (PV) processes.

Despite substantial progress, the global PV market remains dominated by wafer-based silicon (Si) SCs, which account for more than 90%. This dominance is attributed to silicon's near-optimal bandgap for PV conversion, its abundance in Earth's crust, nontoxicity, and the benefits of decades of industrial development in the semiconductor and chemical sectors. Laboratory record efficiencies reached 26% for p-type silicon cells [1-3], enabled by heterojunction (HJ) architecture. Commercial silicon modules have now achieved efficiencies of 24.4%, with both laboratory and industrial values steadily improving. The technological maturity and scalability of silicon PV systems are continuously improving [4]. Nevertheless, the performance of single-junction SCs is inherently limited by the Shockley-Queisser (SQ) limit. A maximum achievable PCE of approximately 33% under realistic solar spectra for band gaps between 1.1 and 1.4 eV [5-6]. Though Si is plentiful and technologically mature, it is inhibited by several inherent limitations. Wafer-based silicon SCs require high-temperature, energy-consuming fabrication procedures, leading to elevated manufacturing costs due to their high melting point [7]. Moreover, Si suffers from weak absorption in the near-infrared spectral area (≈1000-1450 nm), which is roughly $10^{-7}$ $cm^{-1}$ near the band edge [7-8], necessitating thick absorber layers and further increasing material usage and fabrication cost. Consequently, considerable research efforts have been devoted to identifying alternative semiconductor materials capable of delivering high photovoltaic performance through low-cost thin-film technologies.

Thin film solar cells (TFSCs) have emerged as attractive candidates for next-generation photovoltaic devices because they require significantly less semiconductor material, can be fabricated on lightweight and flexible substrates, and offer excellent opportunities for bandgap engineering and heterostructure design. Over the past decades, absorber materials such as CdTe, $Cu(In,Ga)Se_2$ (CIGS), and metal-halide perovskites have achieved remarkable efficiencies [9-13]. ]. In the last few years, the PCE of optimal SCs utilizing CIGS absorbers has reached values of 22-23% in laboratory settings. Even with these remarkable accomplishments, the voltage and current density of those cells only reach roughly 84% and 88% [10]. This suggests that the accumulation of charge carriers requires significant enhancement, and the rate of recombination must also be markedly decreased [11]. Nevertheless, these technologies continue to face important challenges, including the scarcity and high cost of constituent elements, toxicity concerns, long-term operational instability, and complex fabrication procedures [12-13]. These limitations have motivated the exploration of new semiconductor absorbers composed of relatively abundant, stable, and environmentally acceptable elements while maintaining favorable optoelectronic properties for efficient photovoltaic energy conversion.

The ongoing search for progressive optoelectronic and renewable energy materials has brought ternary and quaternary chalcogenides into the limelight, basically due to their suitable electronic bandgap values and excellent ability to absorb the electromagnetic radiation. Recently, the $I_3$-V-$VI_4$ group semiconductors specifically the Ag-based in the cubic form $Ag_3TaX_4$ (X=S, Se, Te) sulvanite structure have emerged as a predominantly potential group of novel p-type semiconductors [14]. The main features of these materials are the unique crystal structure that directly drives the most impressive physical, optical and electronic characteristics [15]. This structure shows the remarkable mechanical and dynamic stability which provide the highly ductile nature which suggest the appropriate the physical properties for the thin-film and long-term photovoltaic device applications [14, 16-17]. Beyond these physical properties, their specific structural system creates an ideal situation for the solar energy conversion. Within the bandgap value of 1.24~1.70 eV range that enables the significant light absorption of order $10^5$ $cm^{-1}$ in the visible range. As a result, $Ag_3TaX_4$ compounds show the large light-absorbing material which are good for the solar cells applications. The Ag-based compounds also exhibit a combination of good electrical conductivity, high Seebeck coefficient, and low lattice thermal conductivity which make them excellent thermoelectric materials [14,18]. The $Ag_3TaX_4$ compounds with a bandgap value close to the optimum value for single-junction photovoltaic conversion which offers particularly for the absorber-layer material. Though the considered material strongly suggest that these are excellent for the PV compounds, their PV performance within complete solar-cell device design and mechanism has not been yet systematically investigated yet.

Motivated by these prospects, the n-CdS/p-$Ag_3TaX_4$/$p^+$-GeS TFSC architecture presents an appealing pathway for next-generation PV technologies. In this proposed structure, noble $Ag_3TaX_4$ has been used as the primary absorber layer, leveraging its favorable optoelectronic properties for efficient light harvesting. A well-established n-type window layer, Cadmium sulfide (CdS), is used to facilitate the efficient charge separation and transportation. The $p^+$-GeS layer, a layered semiconductor with strong anisotropic optical absorption and high hole concentration, acts as an effective back-surface field (BSF) by suppressing rear-interface recombination and enhancing the built-in electric field. The integration of the window, CdS, and the BSF, GeS, with $Ag_3TaX_4$ (X = S, Se, Te) as the primary absorber has not been extensively studied, offering an opportunity to explore a novel class of efficient thin-film heterostructures.

Herein, $Ag_3TaX_4$-based photovoltaic devices have been designed employing CdS as the window and GeS as the BSF layers. CdS is a prominent n-type compound that has received significant interest due to its superior transmission properties, adjustable and extensive optical bandgap, elevated carrier concentration, and remarkable photostability during prolonged light exposure [19]. It is a II-VI compound, which has a band gap of 2.45 eV. CdS strongly absorbs ultraviolet light with photon energies exceeding its band gap. In the meantime, the CdS allows visible and NIR spectra to pass through the window layer. Since the absorbed ultraviolet area constitutes a minor fraction of the total photon spectra, the associated optical loss is insignificant, making CdS an appropriate window layer for both visible and infrared spectral regions [20-21]. There are different ways of depositing CdS. Chemical bath deposition (CBD), a highly versatile and widely adopted deposition method, is one of them. It is cheap, doesn't waste any materials, can be done over and over again, and is simple to set up [22]. On other hand, GeS belongs to the group-IV monochalcogenide family and has a band gap between 1.6 and 1.7 eV [23-24]. It exhibits a range of advantageous optoelectronic properties, including excellent chemical and environmental stability, earth abundance, and the absence of toxic elements. Owing to its tunable bandgap, GeS offers a favorable energy-band arrangement with the $Ag_3TaTe_4$ layer. In addition, it features a light-absorption characteristic exceeding $10^4$ $cm^{-1}$, strong electron affinity, preeminent carrier mobility, and strain- or field-tunable electronic and optical behavior. These attributes render GeS a highly suitable candidate for TFSC devices.

Theoretical modeling and numerical simulation have become an indispensable approach for evaluating emerging photovoltaic materials prior to experimental fabrication. Advanced simulation platforms enable comprehensive investigation of carrier transport, electrostatic potential distribution, optical generation, recombination mechanisms, and interface properties while significantly reducing experimental cost and development time [25]. Among the available simulation tools, COMSOL Multiphysics provides a versatile finite-element framework capable of solving semiconductor transport equations in realistic three-dimensional (3D) geometries. Unlike one-dimensional (1D) simulation approaches, 3D finite-element modeling captures spatial variations of electric fields, carrier concentrations, and recombination processes with greater physical accuracy, thereby providing deeper insight into device operation and optimization. In this research work, the simulation work was performed in the COMSOL Multiphysics software. The semiconductor module in COMSOL is used to figure out the charge transportation for electrons and holes by solving the drift-diffusion and Poisson's equations. Optical generation can be incorporated to study wavelength-dependent absorption in the absorber layer. This enables the prediction of key performance metrics, such as electric field, J-V characteristics, thermal behavior, and the effects of material and design variations [7, 26-27].

This study investigates the design and theoretical performance of a novel n-CdS/p-$Ag_3TaX_4$/$p^+$-GeS TFSC utilizing COMSOL Multiphysics software. Comprehensive experiments have been undertaken to evaluate the effects of varied parameters within each stratum. The device exhibits reasonable PV characteristics. The findings provide valuable insights into optimal device design and highlight the potential of $Ag_3TaTe_4$-based heterostructures for next-generation TFSC technologies.

## 2. $Ag_3TaX_4$ (X = S, Se, Te) cell construction and simulation outline

### 2.1 Structural arrangement of the $Ag_3TaX_4$ (X = S, Se, Te)-based PV devices

Fig. 1(a) illustrates the conceptual plan of the proposed TFSC devices. Here, n-type CdS is incorporated at the top of the devices as the window layer. CdS is preferred due to its highly stable properties and its adjustable and wider bandgap. Also, it possesses superior carrier mobility and high light-transmissive properties [28-30]. The $p^+$-GeS is used as the BSF layer with a bandgap of 1.65 eV. The combination of the absorber layer $Ag_3TaX_4$ (X = S, Se, Te) and the BSF layer GeS forms a p-$p^+$ interface that generates an internal electric field. This field suppresses electron recombination with holes at the rear surface, thereby reducing carrier recombination.

Fig. 1(b) reveals the illumined band structure of the proposed cells. Here, quasi-Fermi levels are between the conduction band, $E_C$, and the valence band, $E_V$. For electrons, the quasi-Fermi level ($E_{Fn}$) is present just below the $E_C$, while for holes ($E_{Fp}$), it is just above the $E_V$. The band alignment is perfected with the carefully chosen materials with desired electron affinity, $E_A$, optical gap, $E_g$, and overall ionization energy, $E_i$. In the CdS/$Ag_3TaX_4$ junction valence band offset (VBO) and the conduction band offset (CBO) show that $E_V$ of $Ag_3TaX_4$ is higher than that of CdS and indicates a large barrier in hole transportation, illustrated in Fig. 1(b). In the $Ag_3TaX_4$/GeS interface, the VBO and CBO indicate that the $E_V$ of the GeS is in a higher position than $Ag_3TaX_4$. This facilitates the structure to transport holes efficiently and prevents the electrons from leaking. A cathode accumulates electrons, which is a metal grid, and a back metal contact acts as an anode to accumulate holes. Silver, Ag, and Nickel, Ni, are used as cathode and anode, correspondingly, to collect charge carriers efficiently.

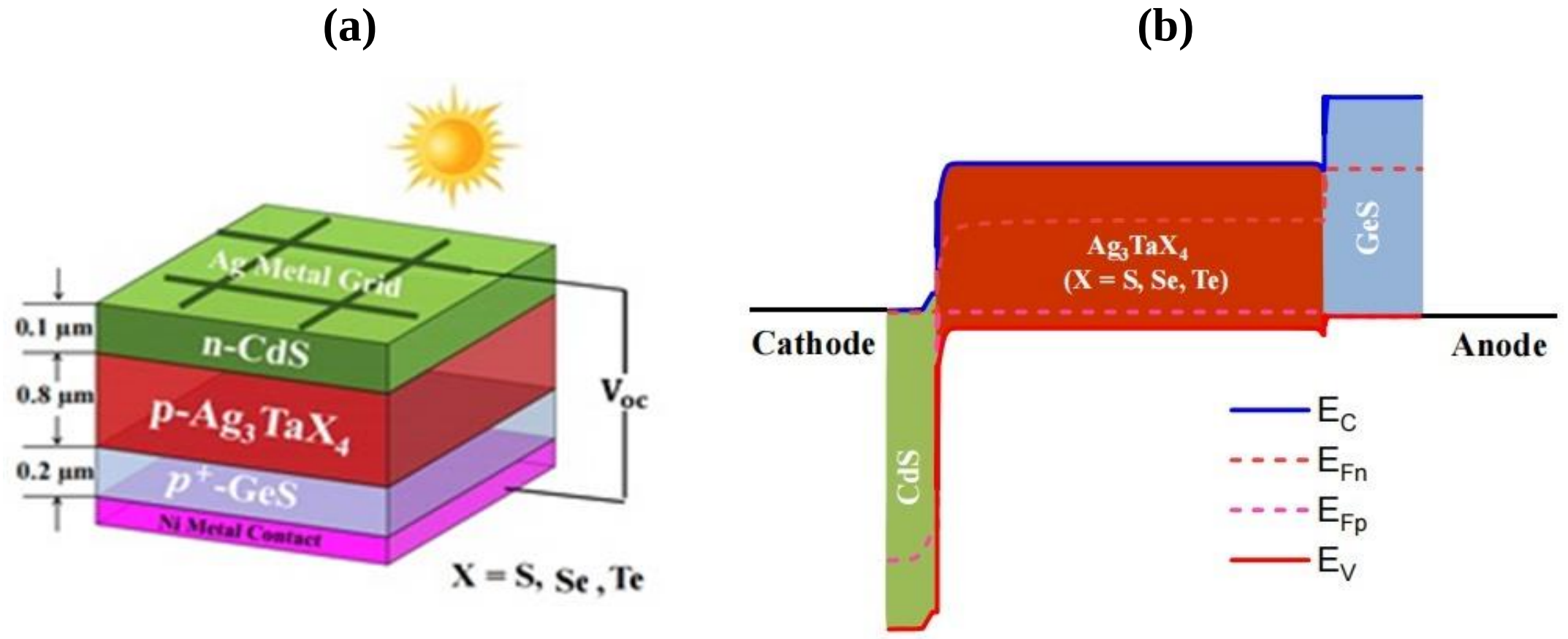

**Fig. 1:** (a) Conceptual 3D diagram and (b) energy band diagram of the n-CdS/p- $Ag_3TaX_4$ (X = S, Se, Te)/$p^+$-GeS thin film solar cells.

### 2.2 Simulation methodology

There are many strategies to simulate and analyze the devices structure. Conventional experimental approaches are often time-consuming and costly as it evaluates the effect of new materials on devices performance, whereas computational techniques provide a faster and more efficient approach for performance optimization. Various software (SCAPS-1D, PC-1D, ATLAS, Silvaco TCAD, Ansys Lumerical, COMSOL Multiphysics) have been used for numerical analysis of solar cells. COMSOL Multiphysics is a well-established option to perform a 3D realistic design and conduct numerical analysis. The proposed SC devices is analyzed using a numerical model implemented in COMSOL, where the Semiconductor Module is used to solve the partial differential equations for charge carrier transport and light-matter interactions. This approach is highly versatile, as it can address a wide range of physical phenomena. The finite element method (FEM) operates by discretizing the devices into a collection of smaller elements, forming a mesh. A boundary value problem is then formulated, generating a system of algebraic equations that approximates the governing functions across the entire domain. Owing to this capability, FEM is widely employed in the simulation and modelling of advanced devices and physical systems [26-27, 31]. The device is designed as a 3D architecture to capture spatial variations. For essential PV processes, this model accounts for the generation of the charge carriers, photon absorption, carrier migration and recombination, and charge extraction. This model is able to give the J-V features with standard sunlight conditions. In this simulation, AM 1.5G incident illumination between 280 nm and 3000 nm is used.

To approximate the absorption edge behavior near the bandgap we use the photo-absorption constant (α) for each layer which is calculated by the square root model (Tauc model) [7,32]. In active layers, carrier generation rate was determined by using [7,33]:

$$G = \eta_0 \frac{P\lambda}{hc} \alpha e^{-\alpha z} \quad (1)$$

Where, G = rate of carrier generation, $\eta_0$ = quantum efficiency, P = incident power, h = Planck's constant, c = light speed, and z = depth into the cell.

The electrical response of the SC devices is defined by the Poisson equation coupled with the continuity equations for electrons and holes. In 3D co-ordinate, it can be defined as [7]:

$$\frac{\delta^2 \Phi}{\delta x^2} + \frac{\delta^2 \Phi}{\delta y^2} + \frac{\delta^2 \Phi}{\delta z^2} = - \frac{\rho}{\varepsilon_0 \varepsilon_r} \quad (2)$$

where the spatial dimensions are represented as x, y, and z of the devices. The Poisson equation, together with the electron and hole continuity equations are solved with this well regarded tools [7,33-34].

In this study, the recombination is calculated using the Shockley-Read-Hall (SRH) model. SRH recombination is given by [7,34-35].

$$R_n = R_p = \frac{np - n_{i,mod}^2}{\tau_p(n+n_1) + \tau_n(p+p_1)} \quad (3)$$

$$n_{i,mod} = \gamma_n \gamma_p \sqrt{N_c N_v} \exp\left(-\frac{E_g \Delta E_g}{2V_{th}}\right) \quad (4)$$

$$n_1 = \gamma_n \sqrt{N_c N_v} \exp\left(-\frac{E_g \Delta E_g}{2V_{th}}\right) exp\left(\frac{\Delta E_t}{V_{th}}\right) \quad (5)$$

$$p_1 = \gamma_p \sqrt{N_c N_v} \exp\left(-\frac{E_g \Delta E_g}{2V_{th}}\right) exp\left(\frac{\Delta E_t}{V_{th}}\right) \quad (6)$$

Where $\gamma_n$ and $\gamma_p$ indicate the degradation constants for electrons and holes, individually; $N_c$ and $N_v$ characterizes the density of states (DOS) in the $E_C$ and the $E_V$, correspondingly; $E_g$ corresponds to the material's energy band gap, while $\Delta E_g$ resembles the band energy reduction. $\tau_n$ and $\tau_p$ are demonstrated as effective lifetimes for electrons and holes, respectively, and $V_{th}$ is the thermal voltage. The $N_v$ and $N_c$ can be determined from the following equations [28]:

$$N_v = 2\left(\frac{m_h^* KT}{2\pi\hbar^2}\right)^{\frac{3}{2}} \quad (7)$$

$$N_c = 2\left(\frac{m_e^* KT}{2\pi\hbar^2}\right)^{\frac{3}{2}} \quad (8)$$

where, $m_h^*$, $m_e^*$, K, T and $\hbar$ stand for the effective masses of hole and electron, Boltzmann constant, temperature and reduced Planck constant, in turn. The hole and electron effective masses of $Ag_3TaX_4$ are taken from the literature [14]. Table 1 depicts the various parameters used in the simulations.

**Table 1:** Simulation parameters of the n-CdS/p-$Ag_3TaX_4$ (X = S, Se, Te)/$p^+$-GeS solar devices.

| **Parameters** | **n-CdS [28]** | **p-$Ag_3TaS_4$ [14]** | **p-$Ag_3TaSe_4$ [14]** | **p-$Ag_3TaTe_4$ [14]** | **$p^+$-GeS [23-24]** |
|---|---|---|---|---|---|
| Thickness (μm) | 0.1 | 0.8 | 0.8 | 0.8 | 0.2 |
| Optical gap (eV) | 2.4 | 1.7 | 1.5 | 1.24 | 1.65 |
| Electron affinity (eV) | 4.4 | 4.2 | 4.2 | 4.2 | 3.5 |
| Relative permittivity | 10 | 10.25 | 10.5 | 14.8 | 9.5 |
| Effective density of states at CB ($cm^{-3}$) | $2.2\times 10^{18}$ | $1.69\times 10^{19}$ | $2.26\times 10^{19}$ | $2.38\times 10^{19}$ | $2.68\times 10^{18}$ |
| Effective density of states at VB ($cm^{-3}$) | $1.8 \times 10^{19}$ | $1.31\times 10^{20}$ | $5.75\times 10^{19}$ | $1.1\times 10^{20}$ | $1.4\times 10^{19}$ |
| Electron mobility ($cm^2\ V^{-1}\ s^{-1}$) | 100 | 87.9 | 72.3 | 70 | 696 |
| Hole mobility ($cm^2\ V^{-1}\ s^{-1}$) | 25 | 17.8 | 30.8 | 20 | 304 |
| Donor density, $N_D$ ($cm^{-3}$) | $1.0\times 10^{18}$ | 0 | 0 | 0 | 0 |
| Acceptor density, $N_A$ ($cm^{-3}$) | 0 | $1.0\times 10^{18}$ | $1.0\times 10^{18}$ | $1.0\times 10^{18}$ | $1.0\times 10^{19}$ |
| Defect volume ($cm^{-3}$) | $1.0\times 10^{14}$ | $1.0\times 10^{14}$ | $1.0\times 10^{14}$ | $1.0\times 10^{14}$ | $1.0\times 10^{14}$ |

### 2.3 Meshing analysis

The computational grid is refined by imposing the element sizes of the design. A lower maximum element size delineates the higher mesh. In this study, 19.8 nm of maximum element size is applied to achieve an extra-fine mesh as shown in Fig. 2. A customized meshing strategy is employed for the numerical modeling throughout the devices with a minimum element size of 0.16 nm. To preserve numerical stability and ensure uniform accuracy throughout all regions, an associated growth and curvature factor is required. The growth rate of 1.08 and a curvature factor of 0.3 are enforced to find the desired mesh structure. Free triangular mesh is applied to the face of the window layer, and then the mesh is swept through the whole devices, producing a highly structured and fine discretization. In particular, symmetry was imposed on all faces of the layers to guarantee homogeneous spatial resolution. The automatic tessellation scheme is configured, while a fixed distribution is used for domain discretization. Mesh quality was assessed using standard indicators, with particular emphasis on skewness and element growth rate. Skewness quantifies the deviation of mesh elements from their ideal geometrical shapes, serving as a key measure of element distortion. The growth rate, on the other hand, characterizes the relative change in element size between adjacent regions. A growth rate of unity corresponds to a perfectly uniform mesh, whereas values below one represent a smooth and gradual transition in element dimensions across the computational domain.

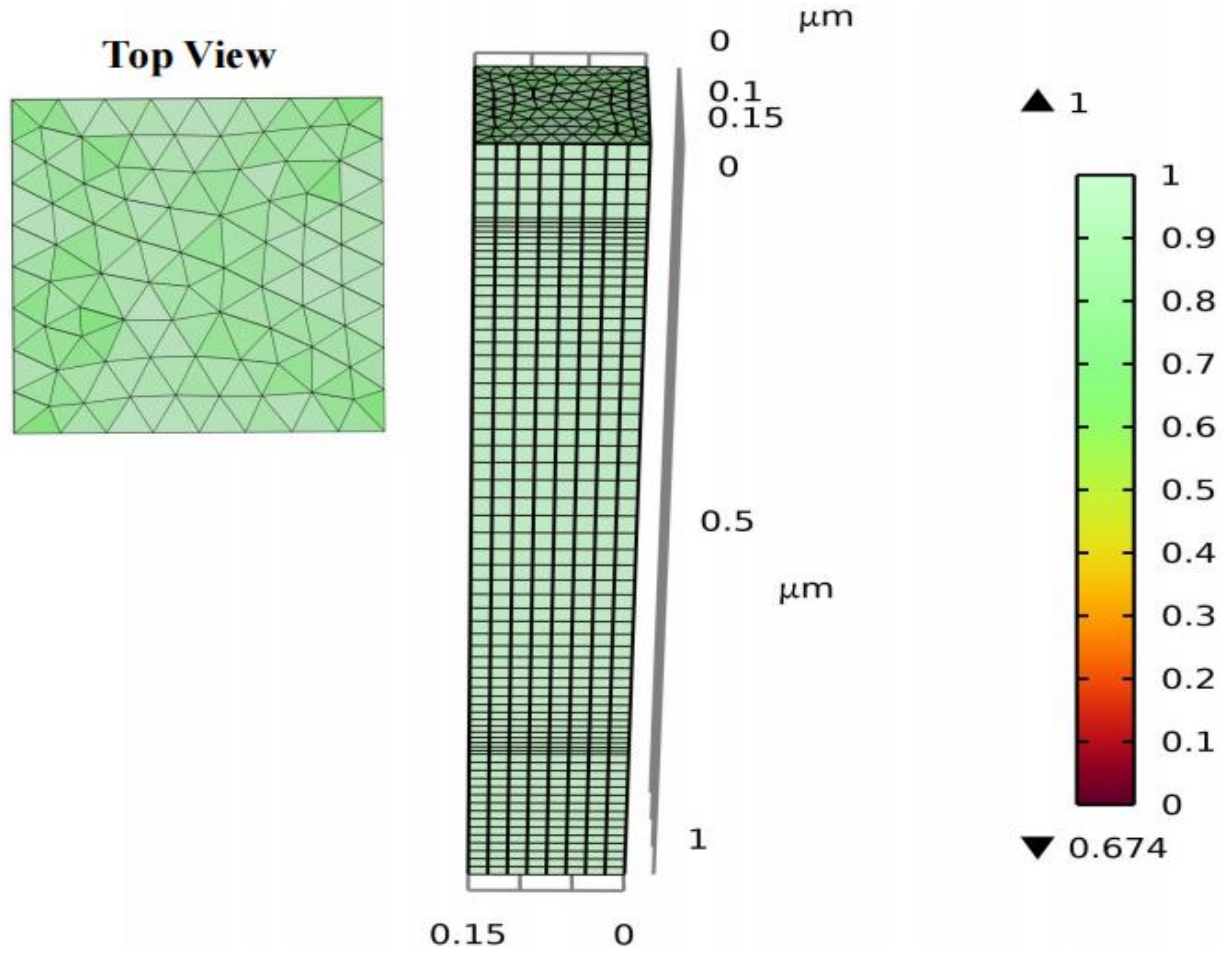

**Fig. 2** Meshing condition of the proposed $Ag_3TaX_4$-based thin film solar cell.

## 3. Results and discussion

### 3.1 Photoinduced carrier generation and electric field profile

Fig. 3(a)-(c) provide a visualisation of the carrier generation and the normalized electric field mapping when the devices voltage is at the open circuit condition for the $Ag_3TaS_4$, $Ag_3TaSe_4$ and $Ag_3TaTe_4$-based SCs. The maximum carrier generation occurs in the vicinity of the n-CdS/p-$Ag_3TaX_4$ junction, which is approximately $5.37\times10^{27}$ cm$^{-3}$ s$^{-1}$, $7.82\times10^{27}$ cm$^{-3}$ s$^{-1}$ and $1.21\times10^{28}$ cm$^{-3}$ s$^{-1}$ for $Ag_3TaS_4$, $Ag_3TaSe_4$ and $Ag_3TaTe_4$, respectively. It happens due to the absorption of incoming photons at the interface. This generation rate gradually decays as one moves deeper into the absorber layer from the surface. In the active region, most of the photons are absorbed; because of this, carrier generation within the BSF region is relatively low. In the absorber and BSF layer interface, carrier generation rate is approximately $2.58\times10^{26}$ cm$^{-3}$ s$^{-1}$, $3.68\times10^{25}$ cm$^{-3}$ s$^{-1}$ and $6.19\times10^{24}$ cm$^{-3}$ s$^{-1}$ of $Ag_3TaS_4$, $Ag_3TaSe_4$ and $Ag_3TaTe_4$, respectively [7,33-36].

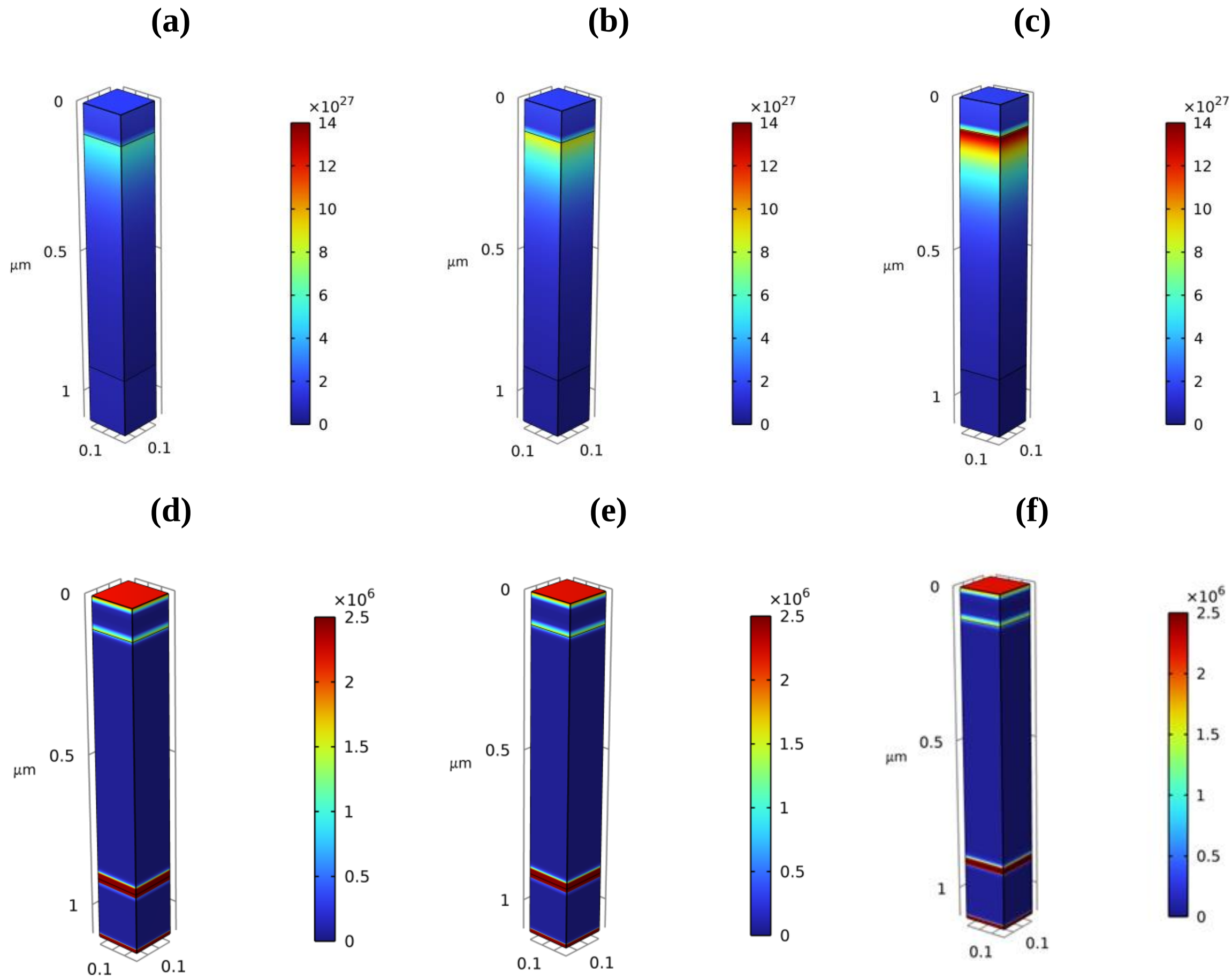

**Fig. 3:** Light induced carrier generation rate mapping of the (a) $Ag_3TaS_4$, (b) $Ag_3TaSe_4$, and (c) $Ag_3TaTe_4$, and Electric field mapping of the (d) $Ag_3TaS_4$, (e) $Ag_3TaSe_4$, and (f) $Ag_3TaTe_4$ -based PV cell in 3D structure.

Fig. 3(d)-(f) also display that the electric field strength in a solar cell is directly related to the band curving at the interfaces, offering a critical understanding of the transport and distribution of charge carriers, ultimately the devices behavior [26]. The gradient of this leads to solving Poisson's equation. Fig. 3 also shows that the electric field distribution is non-uniform with observable peaks placed at the CdS/absorber and absorber/BSF interfaces. These peaks occur due to the build-up of charge carriers in the heavily doped areas of the $Ag_3TaX_4$(X = S, Se, Te) -based solar cell devices.

### 3.2 Altering the physical parameters of the $Ag_3TaX_4$ (X = S, Se, Te) region.

The performance metrics of the $Ag_3TaX_4$ (X = S, Se, Te) based PV devices, when altering the thickness, concentration of doping, and defect density in the absorber layer, are exhibited in Fig. 4.

Fig. 4(a) displays the variation of performance metrics $V_{OC}$, $J_{SC}$, FF, and PCE with the change of breadth in the absorber region. An increase in the breadth of $Ag_3TaS_4$, $Ag_3TaSe_4$ and $Ag_3TaTe_4$ from 0.4 µm to 1.2 µm causes an insignificant fall in the $V_{OC}$ of all absorbers. However, $J_{SC}$ increases with the breadth of $Ag_3TaS_4$, $Ag_3TaSe_4$, and $Ag_3TaTe_4$ absorber layer by 3.39 mA/cm$^2$, 3.85 mA/cm$^2$, and 3.44 mA/cm$^2$, consequently. As the depth of the absorbing layer increases, the photon absorption thereby improves electron-hole generation, resulting in $J_{SC}$ increasing between the range of 0.4 µm and 1.2 µm [36]. FF also slightly increases from 86.66% to 88.00% for $Ag_3TaSe_4$ and 86.21% to 86.54% for $Ag_3TaTe_4$. However, a downward trend takes place for $Ag_3TaS_4$, and FF slightly falls from 86.12% to 85.46%. The impact of the absorber layer PCE directly depends on the combined contribution of $V_{OC}$, $J_{SC}$, and FF. Hence, all of these PV parameters help to improve the resultant PCE increased for $Ag_3TaS_4$, $Ag_3TaSe_4$ and $Ag_3TaTe_4$ from an initial 22.2%, 25.15% and 25.7% to a final 25.08%, 28.41%, and 27.81%, respectively [37].

Fig. 4(b) illustrates the effect of the absorber layer's doping concentration on the performance metrics. An increase in the concentration of doping of $Ag_3TaX_4$ (X = S, Se, Te) from $10^{15}$ to $10^{19}$ cm$^{-3}$ upsurges $V_{OC}$. This may happen due to a downward trend in the dark current $I_0$ with a higher concentration of doping, thereby leading to a rise in $V_{OC}$, as evident from equations (7) and (8) given below [37]:

$$I_0 = An_i^2\left(\frac{D_e}{L_e N_A} + \frac{D_h}{L_h N_D}\right) \quad (9)$$

$$V_{OC} = \frac{kT}{q}\ln\left(\frac{I_L}{I_0} + 1\right) \quad (10)$$

Where A symbolizes the cross-sectional range of the p-n HJ, $n_i$ is the intrinsic carrier concentration, $L_e$ is the diffusion length of minority charges (electrons) in the p-region, $L_h$ is the diffusion length of minority charges (holes) in the n-region, $D_e$ is the diffusion coefficient of electrons, $D_h$ is the diffusion coefficient of the hole, $N_A$ is the donor concentration and $N_D$ is the acceptor concentration. But $J_{SC}$ remains almost constant with the doping concentration.

FF exhibits a significant dependence on the absorber doping concentration, increasing as the acceptor concentration rises from $10^{15}$ cm$^{-3}$ to $10^{19}$ cm$^{-3}$ across all absorbers. Specifically, the FF increases from 66.82% to 89.36% for $Ag_3TaS_4$, from 76.48% to 89.47% for $Ag_3TaSe_4$, and from 77.31% to 86.74% for $Ag_3TaTe_4$. This improvement can be attributed to the enhanced electrical conductivity of the absorber layer at higher doping concentrations, which reduces the parasitic resistance and facilitates more efficient charge transport [38].

Here, PCE growing up with increasing concentration of doping due to combined effect of $V_{OC}$, $J_{SC}$, and FF for all absorber layer materials.

Fig. 4(c) depicts the variation in defect concentration ($10^{12}$ cm$^{-3}$ - $10^{16}$ cm$^{-3}$) of the absorber layer at an optimized thickness of 800nm and an optimized concentration of doping of $10^{18}$ cm$^{-3}$ and its impact on the devices. The $V_{OC}$ deteriorates from 1.57V to 1.24V, 1.35V to 1.04V and 1.04V to 0.76V for $Ag_3TaS_4$, $Ag_3TaSe_4$ and $Ag_3TaTe_4$, consequently, while PCE precipitously declines from 28.23% to 19.81%, from 31.83% to 21.36%, and from 32.04% to 22.23%, respectively.

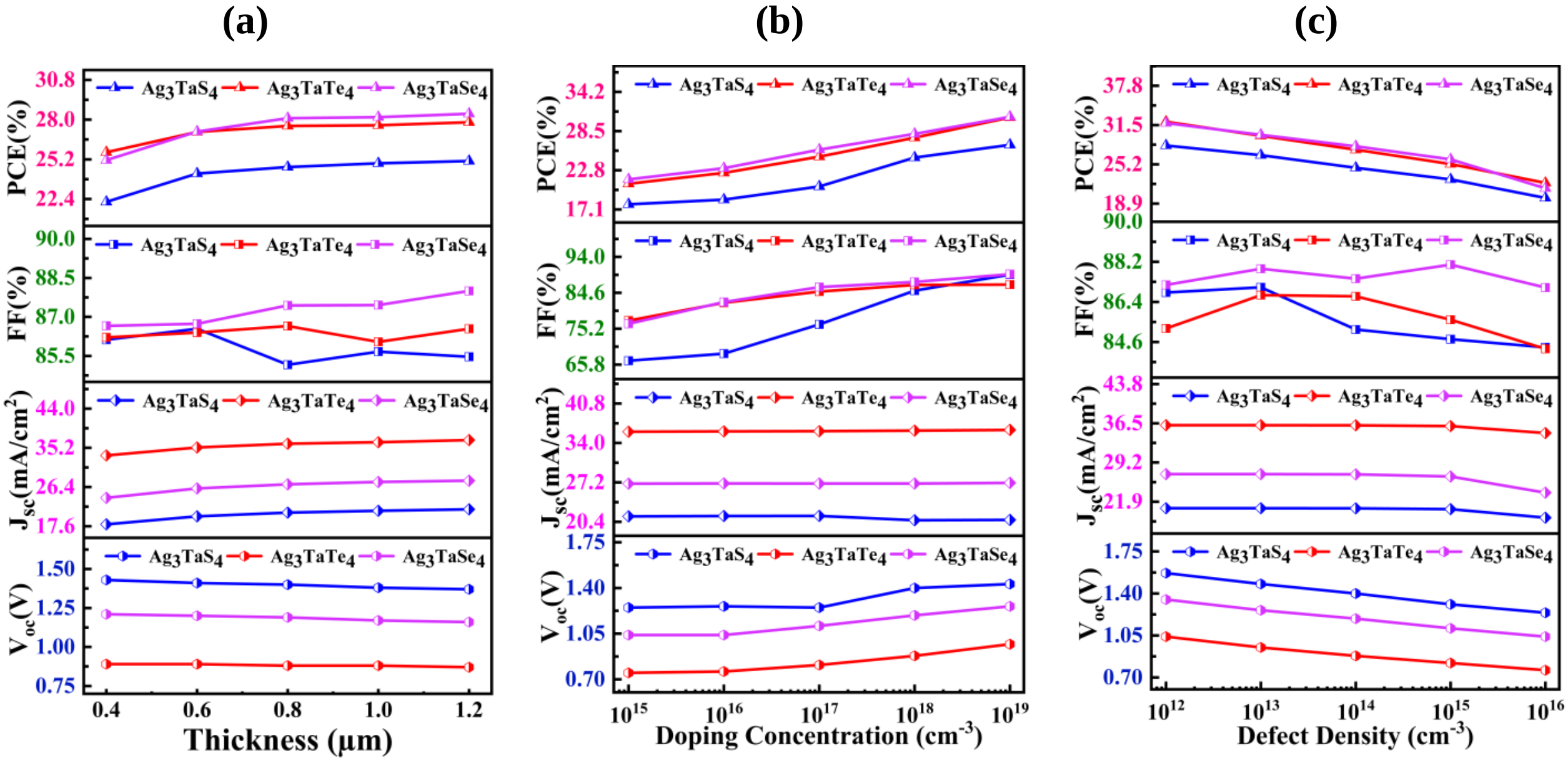

**Fig. 4:** Diversification of (a) Breadth (b) Doping concentration and (c) Defect density in absorber layer of the n-CdS/ p-$Ag_3TaX_4$ (X = S, Se, Te)/$p^+$-GeS hetero junction Solar cell.

Also observed in Fig. 4(c), $J_{SC}$ is almost constant for the variation from $10^{12}$ cm$^{-3}$ to $10^{15}$ cm$^{-3}$. However, further increasing the defect parameter up to $10^{16}$ cm$^{-3}$, then $J_{SC}$ decreases. It may belong to the increase in recombination with defect density. The FF also exhibits a declining trend, decreasing as defect density increases. At lower defect concentrations, FF shows a slight improvement when the defect density rises from $10^{12}$ to $10^{13}$ cm$^{-3}$, followed by a gradual degradation at higher defect levels. This may occur because of an elevation in diode ideality factor [28]. These parameters facilitate decreasing the device's overall efficiency by limiting the availability of charge carriers for producing an electric current. Also, defects can produce trap states within the absorber layer materials' bandgap [39-40].

### 3.3 Tuning the physical parameters of the CdS window

Fig. 5 depicts the effect of changing the thickness, concentration of carrier, and defect density of the window region on performance metrics for every three devices.

Fig. 5(a) illustrates the diversity of the performance indicators with changing the thickness of the window layer. $V_{OC}$ is almost constant with changing the width of the window layer between 0.1μm and 0.5 μm. In these PV devices, the $J_{SC}$ gradually decreases as the window layer thickness increases, primarily due to parasitic light absorption and increased series resistance [41]. FF increases in $Ag_3TaS_4$ and $Ag_3TaSe_4$ at dilute thickness, after which it remains nearly unchanged over the thickness range of 0.2–0.5 μm. Subsequently, PCE drops with the thickness diversification of the window layer by 3.13% for $Ag_3TaS_4$, and 2.69% for $Ag_3TaSe_4$, and 2.03% for $Ag_3TaTe_4$. This drop mainly occurred due to parasitic light absorption, which restricts the number of photons reaching the primary absorber layer and increases internal charge recombination [42].

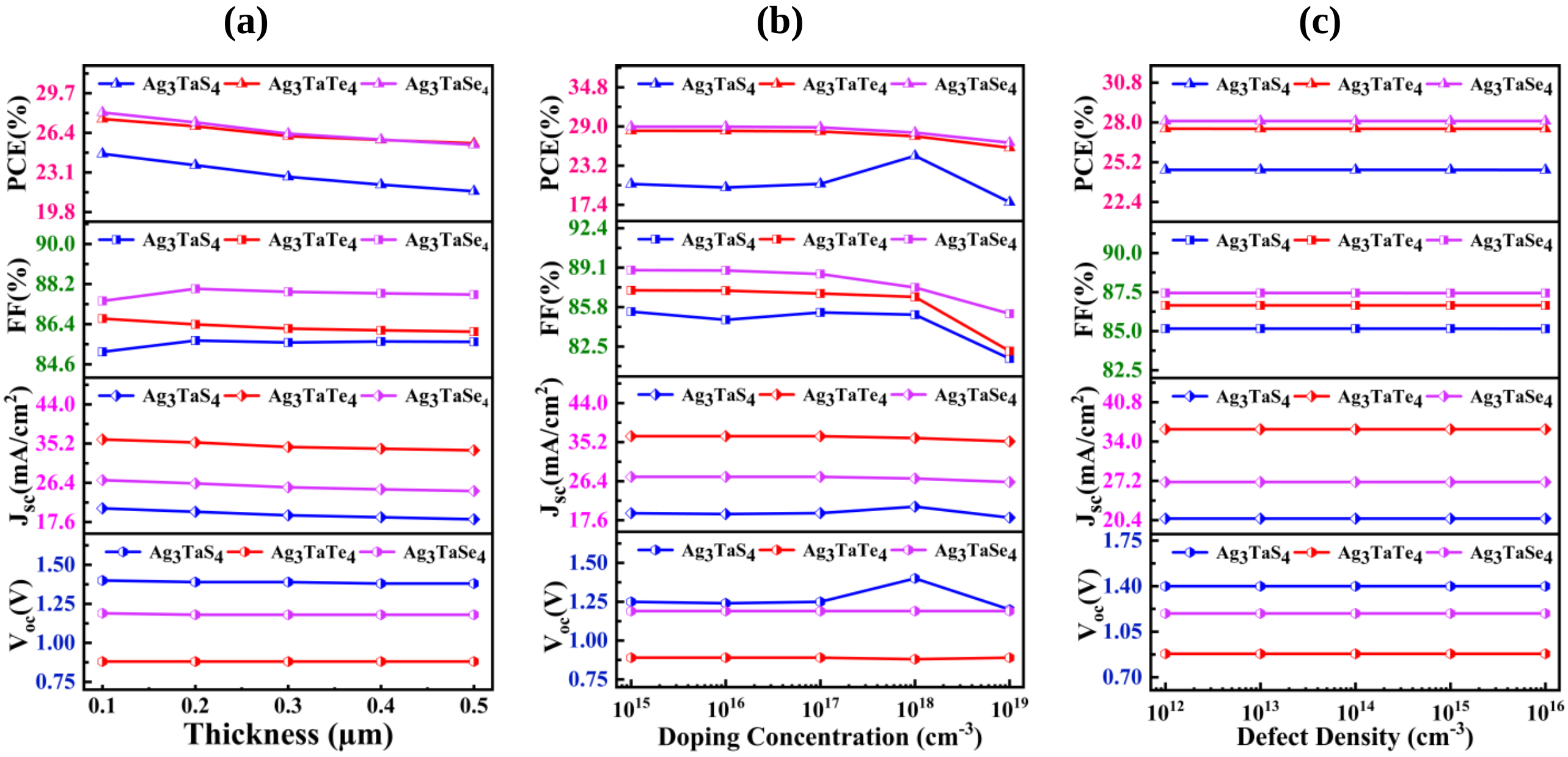

**Fig. 5:** The outcome of changing the (a) thickness, (b) doping, and (c) defect density in the CdS window of the $Ag_3TaX_4$ (X = S, Se, Te)-based heterojunction SCs.

Fig. 5(b) displays that changing the carrier density from $10^{15}$ $cm^{-3}$ to $10^{19}$ $cm^{-3}$ of the buffer layer. $V_{OC}$ is steady at 1.19V for $Ag_3TaSe_4$ and at 0.88V for $Ag_3TaTe_4$. However, $Ag_3TaS_4$ illustrates a steady $V_{OC}$ at 1.25V for a variation of doping from $10^{15}$ $cm^{-3}$ to $10^{17}$ $cm^{-3}$. When further increasing the doping to $10^{18}$ $cm^{-3}$, $V_{OC}$ increases to 1.4V. When the doping density surpasses $10^{18}$ $cm^{-3,}$ then $V_{OC}$ falls from 1.4V to 1.25V due to a negative impact on the material's quality and an increase in recombination. Elevated doping levels can introduce deep-level traps, leading to higher recombination losses [43]. Change of $J_{SC}$ with window layer doping concentration ($10^{15}$ $cm^{-3}$ to $10^{17}$ $cm^{-3}$) is almost steady. However, when the doping concentration sits at $10^{18}$ $cm^{-3}$, Jsc of $Ag_3TaS_4$ is slightly increased. But for $Ag_3TaSe_4$ and $Ag_3TaTe_4$, $J_{SC}$ is teeny drops at higher doping density. An increase in doping concentration drives the n-type quasi-Fermi level at the CdS/$Ag_3TaX_4$ interface closer to the conduction band, while simultaneously displacing the p-type quasi-Fermi level away from the valence band, thereby intensifying the valence-band bending. Due to the increase in the bending of the valence band, recombination increases in CdS, $Ag_3TaX_4$, and CdS/$Ag_3TaX_4$ junctions. $J_{SC}$ decreases [28]. Excessive band bending promotes carrier recombination within the CdS layer, the $Ag_3TaX_4$ absorber, and particularly at the CdS/$Ag_3TaX_4$ interface, which suppresses carrier collection and leads to a reduction in $J_{SC}$ at higher doping concentrations [28]. The FF is steady around 85.0%, 88.0%, and 86.0% correspond to $Ag_3TaS_4$, $Ag_3TaSe_4$, and $Ag_3TaTe_4$, respectively. However, a further increase in impurity to $10^{19}$ $cm^{-3}$ establishes a decay in the FF. The elevation of the impurity concentration from $10^{15}$ $cm^{-3}$ up to $10^{19}$ $cm^{-3}$ drops the PCE. In $Ag_3TaSe_4$, it is around 2.38% and for $Ag_3TaTe_4$, it is around 2.5%. Although in $Ag_3TaS_4$, the PCE is increased to 24.66% at $10^{18}$ $cm^{-3}$ and then drops to 17.79% at $10^{19}$ $cm^{-3}$. This performance degradation may originate from increased free-carrier recombination at elevated donor concentrations [7, 44-45].

Fig.5(c) demonstrates that the variation of defect ranges from $10^{12}$ $cm^{-3}$ to $10^{16}$ $cm^{-3}$ in the CdS window regions does not significantly affect the characteristics of the proposed SC devices, which matches with our recent studies [28,40,46-47].

### 3.4 Investigating the effect of parameters of the BSF

Fig. 6 reveals the impact of changing the physical attributes of GeS BSF on the $Ag_3TaS_4$-, $Ag_3TaSe_4$-, and $Ag_3TaTe_4$-based TFSC devices' performance.

Fig. 6(a) examines the effect on operational indicators for changes in the GeS layer depth ranging from 0.1µm to 0.5 µm. $V_{OC}$ remains almost constant at around 1.4 V, 1.19 V, and 0.88 V for $Ag_3TaS_4$, $Ag_3TaSe_4$, and $Ag_3TaTe_4$, correspondingly, due to this variation. Likewise, the $J_{SC}$ exhibits only insignificant variation over the investigated thickness range. Whereas in $Ag_3TaS_4$, FF is significantly increased from 77.27% to 89.4% with the thickness of GeS. In contrast, Fig. 6(a) also describes that for both $Ag_3TaSe_4$ and $Ag_3TaTe_4$, the Fill factor remains almost unchanged, which is around 87%, and 86% respectively. However, PCE is increased by 3.6% for $Ag_3TaS_4$, 1.25% for $Ag_3TaSe_4$ and 0.68% for $Ag_3TaTe_4$ as the thickness changes from 0.1μm to 0.5 μm because the back surface field layer provides an electric field at the p-p$^+$ interface, thereby preventing minority charges and reducing recombination at the rear contact. That improves the net PCE of the SC devices [44].

Fig. 6(b) tracks the consequence of variation in doping density ($10^{15}$ cm$^{-3}$ - $10^{19}$ cm$^{-3}$) for BSF layer on the devices. The effect of altering the carrier concentrations, ranging from $10^{15}$ cm$^{-3}$ to $10^{18}$ cm$^{-3}$ on $V_{OC}$ is insignificant for all materials. But when the doping density at $10^{19}$ cm$^{-3}$ $V_{OC}$ is increased to 1.4V from 1.34V for $Ag_3TaS_4$ and 1.19V from 1.15V for $Ag_3TaSe_4$. Nevertheless, Impurity has no noticeable impact on $J_{SC}$. Here, the Fill factor remains almost unchanged, which is around 86% for $Ag_3TaTe_4$. At low doping density, FF of $Ag_3TaS_4$ is increased, and it is almost constant for the range of $10^{16}$ cm$^{-3}$ - $10^{18}$ cm$^{-3}$, and then again FF decreased at $10^{19}$ cm$^{-3}$ doping density. On the other hand, FF of $Ag_3TaSe_4$ is fixed in the range of $10^{15}$ cm$^{-3}$ - $10^{18}$ cm$^{-3}$ and it decreased to $10^{19}$ cm$^{-3}$. The PCE of all materials-based SC is increased with concentration doping owing to the strengthening of the built-in electric field at the back interface [48].

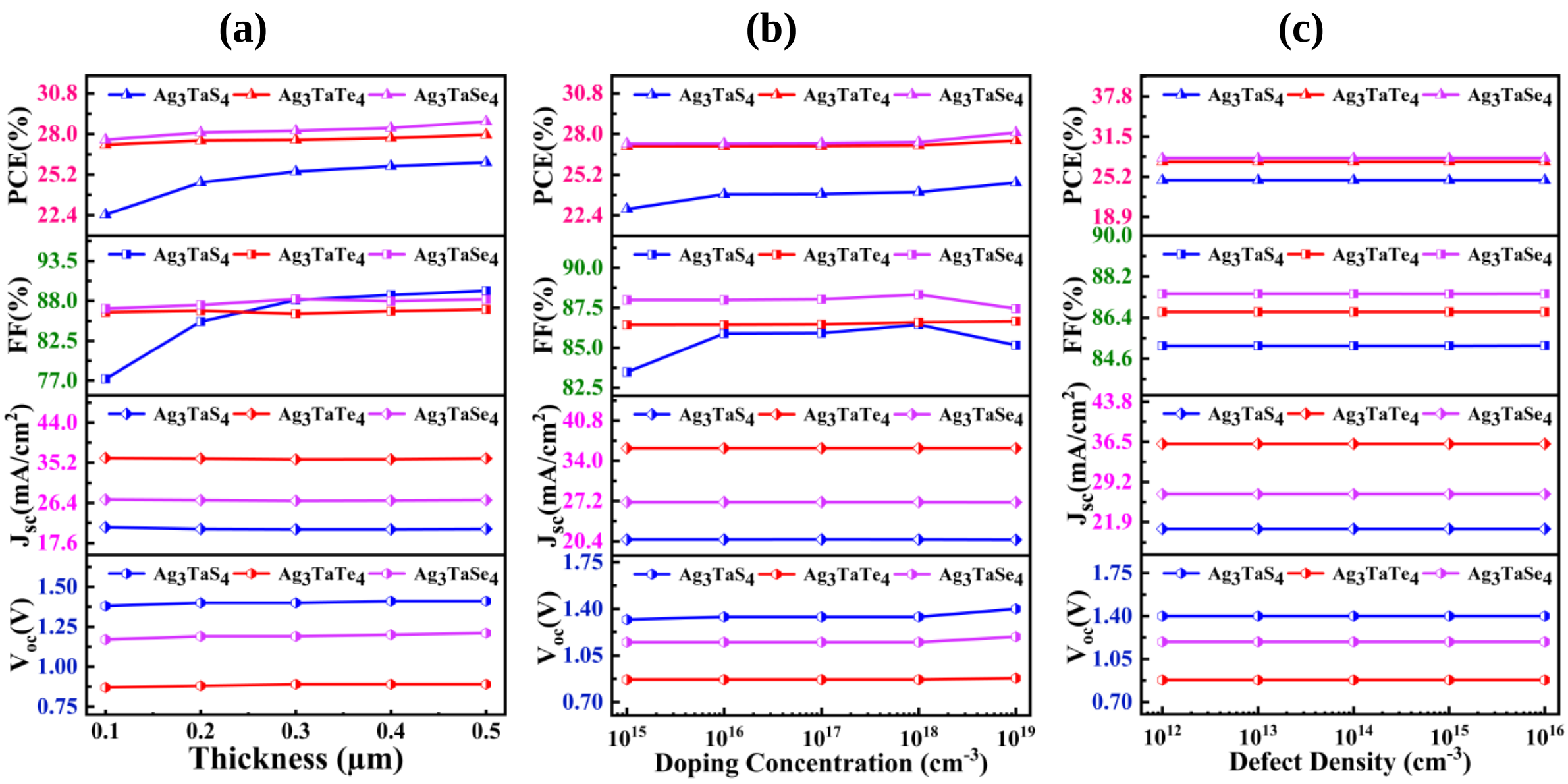

**Fig. 6:** The result of changing the (a) Breadth, (b) Carrier density, and (c) defect density in the BSF region of the $Ag_3TaX_4$ (X = S, Se, Te)-based hetero junction SCs.

Fig. 6(c) demonstrates the effect of the GeS BSF layer defect density on the PV performance of the proposed SCs. As the defect density increases from $10^{12}$ to $10^{16}$, the device characteristics exhibit negligible variation, indicating that the performance is largely insensitive to defects within the layer over the investigated range. This behavior is consistent with our previous findings and other reported studies, which have shown that moderate defect densities in the BSF layer have a minimal impact on the overall photovoltaic performance because most photogeneration and carrier transport occur within the absorber layer [28,40,46,47].

### 3.6 Influence of resistance on the targeted devices

Fig. 7 depicts the influential properties of series resistance ($R_S$) and shunt resistance ($R_{Sh}$) on the targeted SC devices. SC devices' efficiency is strongly influenced by resistive losses, where ideal operation requires zero $R_S$ and infinite $R_{Sh}$. In a practical device, $R_S$ arises from the bulk material, interfaces, and electrical contacts between different layers, while $R_{Sh}$ is governed by recombination pathways, defects, and interface quality [43].

The impact of $R_S$ on device performance is illustrated in Fig. 7(a). As $R_S$ increases from 0 to 10 $\Omega\cdot cm^2$, the $V_{OC}$ and $J_{SC}$ remain stable for all three materials, $Ag_3TaS_4$ (1.4 V, 20.68 $mA/cm^2$), $Ag_3TaSe_4$ (1.19 V, 27.00 $mA/cm^2$) and $Ag_3TaTe_4$ (0.88 V, 36.14 $mA/cm^2$). In contrast, the FF experiences a substantial drop from 85.14% to 72.03% for $Ag_3TaS_4$, 87.43% to 66.68% for $Ag_3TaSe_4$, and 85.82% to 50.53% for $Ag_3TaTe_4$. Consequently, the PCE declines sharply, falling from 24.66% to 20.84%, 28.1% to 21.4%, and 27.56% to 16.19%, respectively. These results underscore that the PCE is highly sensitive to FF degradation, which is the primary driver of efficiency loss as $R_S$ increases [45].

Fig. 7(b) clarifies the influence of $R_{Sh}$ on the PV characteristics of the CPS solar cell, with $R_{Sh}$ varying from $1\times10^1$ to $1\times10^{10}$ $\Omega\cdot cm^2$. For $R_{Sh}$ values less than 1 $k\Omega\cdot cm^2$, the values of PV parameters $V_{OC}$, FF, and PCE are very low. Because of low shunt resistance, the SC creates alternative pathways for light-generated current to bypass [49]. Fig. 7(b) indicates that the device's performance is relatively insensitive to $R_{Sh}$ beyond a certain threshold (1 $k\Omega\cdot cm^2$). For $R_{Sh}$ values exceeding 1 $k\Omega\cdot cm^2$, the variation in PCE is limited to around 0.6%, suggesting minimal sensitivity of the devices toward $R_{Sh}$. Overall, the analysis confirms that high shunt resistance combined with low series resistance is essential for achieving optimal device performance [46].

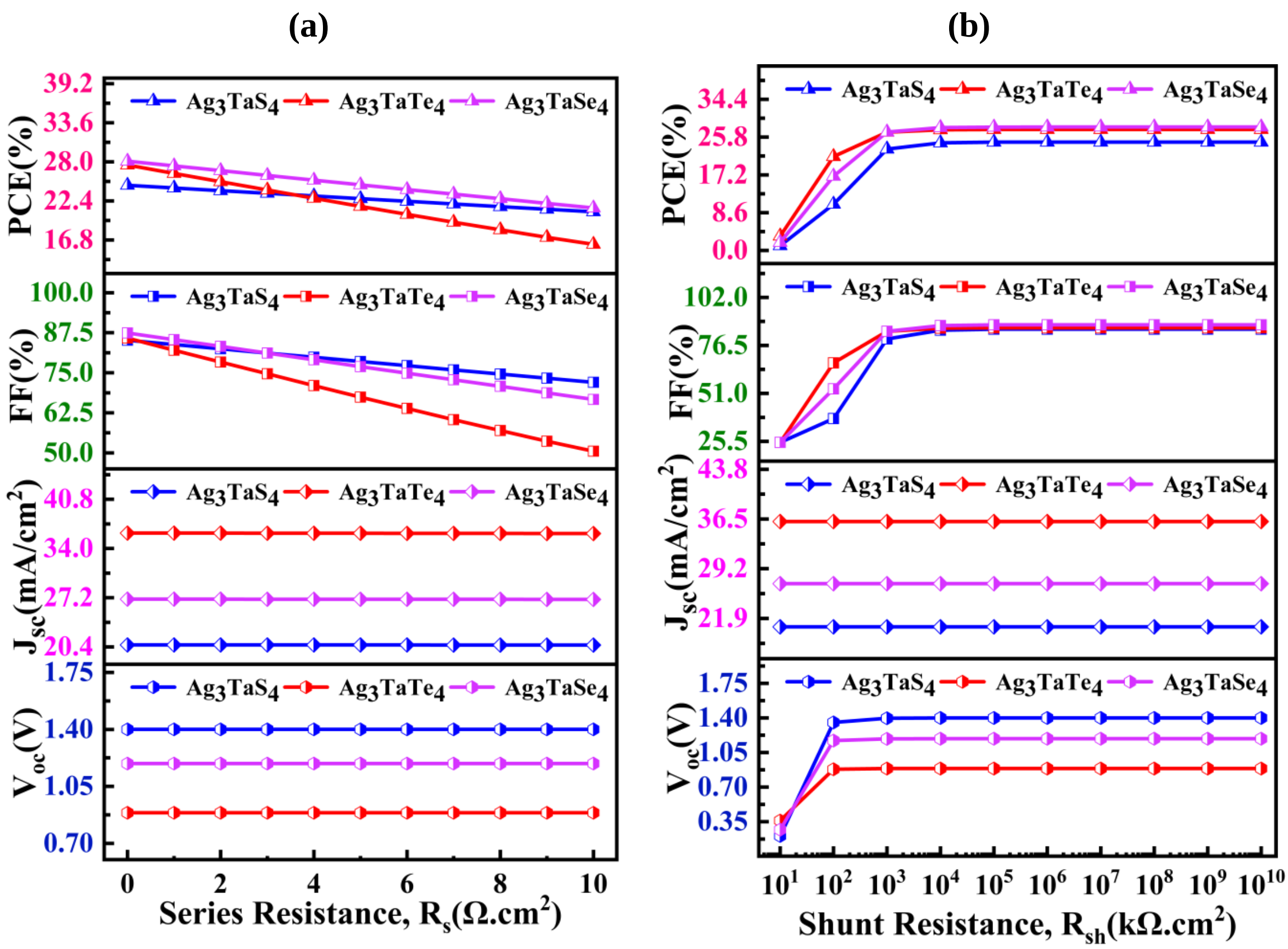

**Fig. 7**: Influence of resistance on the key parameters of the specified $Ag_3TaX_4$ solar devices.

### 3.7 Thermal characteristics of the target devices

#### 3.7.1 Impact of temperature variance on the specified devices

For accurate and proper analysis of solar cells, it is not enough to deal only with electrical and optical characteristics. The thermal effects on solar cells must also be considered. This affects the performance parameters, stability, and long-term durability of the solar cell device.

The influence of temperature across the 200 K to 600 K range on $V_{OC}$, $J_{SC}$, FF, and PCE is presented in Fig. 8. As the temperature increases, all three devices exhibit a pronounced degradation in overall performance. In $Ag_3TaS_4$ and $Ag_3TaSe_4$, $V_{OC}$ reduces by 0.67V and 0.52V, while $J_{SC}$ increases by 0.86 mA/cm$^2$ and 0.45 mA/cm$^2$ with increasing temperature, respectively. The $J_{SC}$ increases with rising temperature due to the bandgap energy of the materials narrows. This narrowing allows the material to absorb a broader spectrum of light to slightly increase $J_{SC}$ [50]. For $Ag_3TaS_4$, FF is minimum at 200K temperature, which is 73.15%, and maximum at 300K temperature, which is 85.16%. Exceeding a temperature of 300 K, FF decays to 78.0%. This variation of FF may occur due to the FF function of the resistance parameters ($R_S$ and $R_{Sh}$) [51]. Similarly, PCE is maximum at 300 K, which is 24.66%, and then PCE drops from 24.66% to 14.72%. For $Ag_3TaSe_4$, a sharp decline in FF (from 85.64% to 76.37%) led to a dramatic collapse in PCE, which plummeted from 30.16% down to 16.49% For the $Ag_3TaTe_4$-based device, $V_{OC}$ demonstrates a significant decrease over the range of 1.02 V to 0.49 V, and $J_{SC}$ rises marginally from 36.04 mA/cm$^2$ to 36.25 mA/cm$^2$ with increasing temperature. FF decays from 90.14% to 67.62%, and PCE significantly drops from 33.14% to 12.01%. This reduced thermal performance is attributed to increased thermal excitation at higher temperatures. Higher temperatures accelerate the carrier recombination within the bulk material, preventing carriers from reaching the depletion region and contributing to current generation. Additional factors influencing the high-temperature efficiency are carrier mobility, hole concentration, and temperature-dependent bandgap narrowing [52].

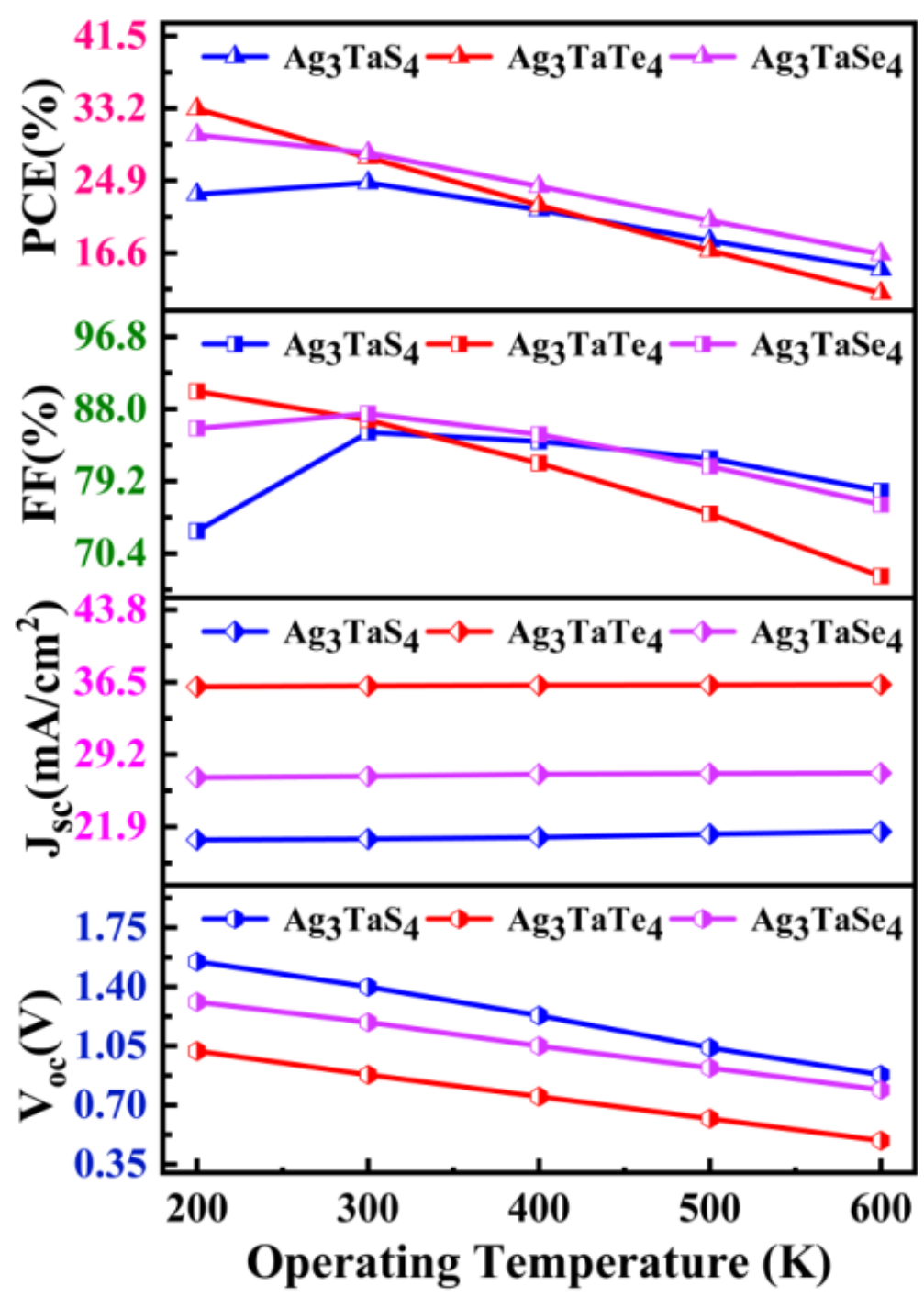

**Fig. 8**: Influence of temperature on the key parameters of the specified $Ag_3TaX_4$ (X = S, Se, Te) photovoltaic devices.

### 3.6.2 Thermal mapping of the $Ag_3TaX_4$ (X = S, Se, Te) devices

#### 3.6.2.1 The 3D Distribution of Joule heating in the proposed devices

Fig. 9 describes the heat distribution due to Joule heating from resistive regions of the $Ag_3TaX_4$ (X = S, Se, Te) SC at the V = 0 V and V = $V_{OC}$ conditions, which is achieved by COMSOL Multiphysics. Joule heating or ohmic heating occurs when an electric current flows through a material that has electric resistance. The moving electrons collide with the atomic lattice, transferring kinetic energy to vibrational energy, thereby producing heat. Consequently, the Joule heat is mainly localized around the contact and junction regions, and the absorber bulk has comparatively low thermal output [27,33]. The Joule-heating distribution of $Ag_3TaS_4$, $Ag_3TaSe_4$, and $Ag_3TaTe_4$ is strongly non-uniform as shown in Fig. 9(a)-(c), respectively, for zero-bias conditions. The maximum heat generation occurs at the CdS/$Ag_3TaX_4$ interface with a value of $9.5\times10^9$ W/m$^3$ in $Ag_3TaS_4$, $1.22\times10^{10}$ W/m$^3$ in $Ag_3TaSe_4$, and $1.5\times10^{10}$ W/m$^3$ in $Ag_3TaTe_4$ then at the BSF surface where the heat-generation intensity value $1.93 \times 10^9$ W/m$^3$ in $Ag_3TaS_4$, $2.93 \times 10^9$ W/m$^3$ in $Ag_3TaSe_4$, and $3.86\times10^9$ W/m$^3$ in $Ag_3TaTe_4$. In comparison, the Joule-heating values of $4.56\times10^8$ W/m$^3$, $5.96\times10^8$ W/m$^3$, and $7.98\times10^8$ W/m$^3$ are observed at the window surface in $Ag_3TaS_4$, $Ag_3TaSe_4$ and $Ag_3TaSe_4$ materials, respectively. The spatial distribution of local Joule heating is confined to the front and rear hetero interfaces, suggesting that the resistive power dissipation occurs mainly in the regions of high local electric field, transport resistance and current crowding. Meanwhile, the negligible temperature rise within the absorber bulk confirms that this layer does not contribute to resistive dissipation, confirming that the resistive losses are mainly related to interface and contact regions instead of the active absorber layer [27,33].

As depicts by Fig. 9(d)-(f), the total Joule-heating intensity is strongly suppressed at V = $V_{OC}$. The maximal heat generation at the CdS/$Ag_3TaX_4$ interface goes down from $9.5\times10^9$ W/m$^3$ to $2.1\times10^9$ W/m$^3$ in $Ag_3TaS_4$, $1.22\times10^{10}$ W/m$^3$ to $2.36 \times 10^6$ W/m$^3$ in $Ag_3TaSe_4$, and $1.5\times10^{10}$ W/m$^3$ to $1.56\times10^8$ W/m$^3$ in $Ag_3TaTe_4$, and at the BSF interface from $1.93\times10^9$ W/m$^3$ to $1.4\times10^7$ W/m$^3$ in $Ag_3TaS_4$, $2.93\times10^9$ W/m$^3$ to $9.75\times10^6$ W/m$^3$ in $Ag_3TaSe_4$, and $3.86\times10^9$ W/m$^3$ to $9.98\times10^8$ W/m$^3$ in $Ag_3TaTe_4$. Similarly, the Joule-heating values at the window surfaces decrease to $2.98\times10^6$ W/m$^3$, $3.38\times10^6$ W/m$^3$, and $2.06\times10^8$ W/m$^3$ for $Ag_3TaS_4$, $Ag_3TaSe_4$, and $Ag_3TaTe_4$, respectively. At the V = $V_{OC}$, the Joule-heating value is reduced significantly. This reduction displays that the power dissipated resistively is much lower at the open-circuit voltage than at the short-circuit condition due to lower current flow through resistive paths. This effectively suppresses localized Joule-heating losses, resulting in superior electrothermal performance and significantly improved operational stability in the optimized device [27,33].

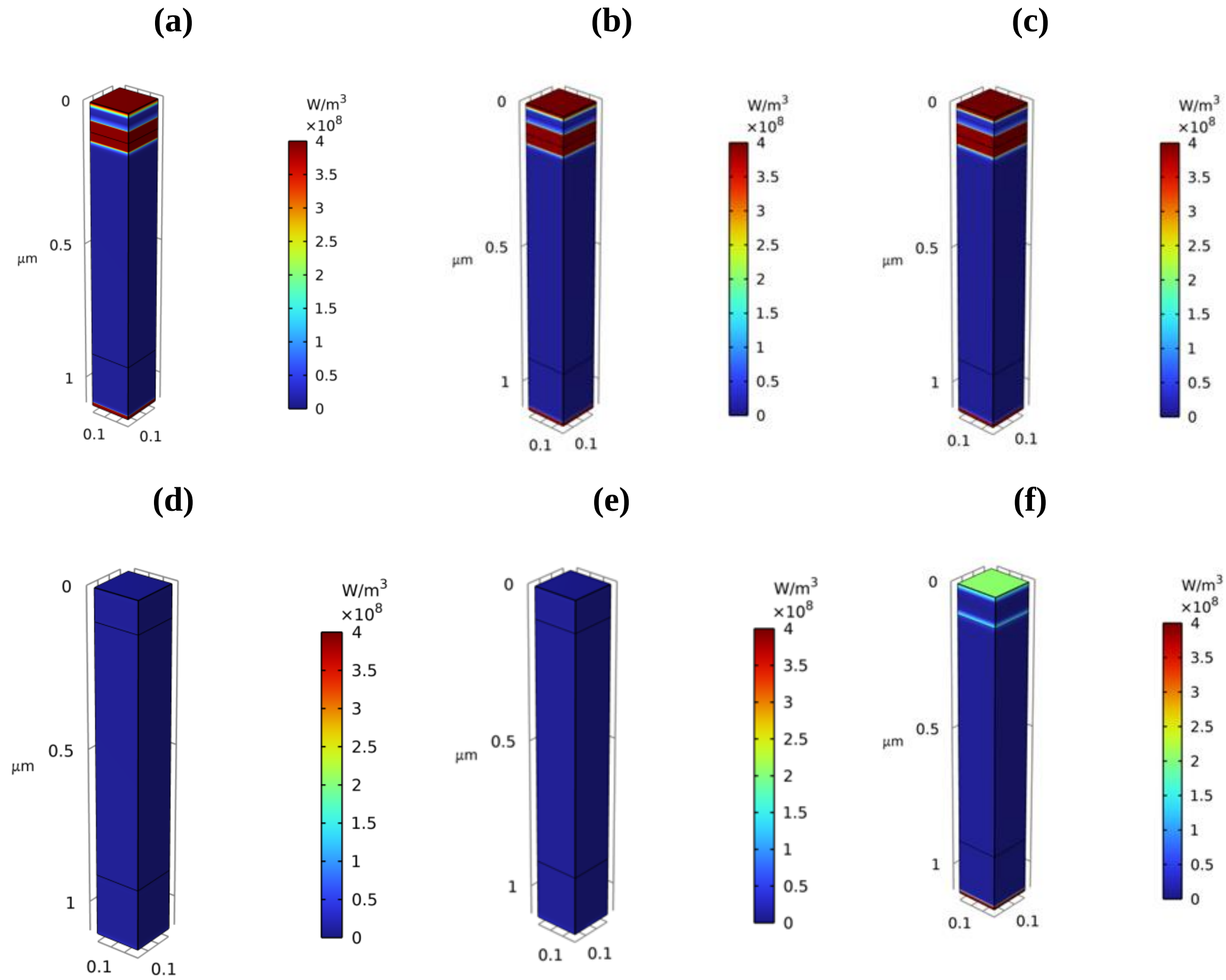

**Fig. 9:** 3D structure of Joule heating at V = 0 V of the (a) $Ag_3TaS_4$ (b) $Ag_3TaSe_4$ and (c) $Ag_3TaTe_4$ and at V = $V_{OC}$ of the (d) $Ag_3TaS_4$ (e) $Ag_3TaSe_4$ and (f) $Ag_3TaTe_4$-based solar cell devices.

### 3.6.2.2 Distribution of Nonradiative Recombination Heating of the proposed devices

Fig. 10 displays the spatial structure of the heat generation from the non-radiative recombination at the V = 0 V and V = $V_{OC}$ conditions. Sunlight generates electron-hole pairs or charges, which are separated and collected to produce electricity. However, they can also recombine. If they recombine radiatively, they emit a photon. If they recombine non-radiatively, the energy is released as heat. Fig. 10(a)-(c) show the heat generation from the non-radiative recombination at V = 0 V for $Ag_3TaS_4$, $Ag_3TaSe_4$ and $Ag_3TaTe_4$, correspondingly. The specific non-radiative recombination heat component is lower at the V = 0 V (short-circuit) condition because most photo generated electrons and holes are extracted out of the cell as electric current before they ever get a chance to recombine [ 53].

As depicts by Fig. 10(d)-(f) at V = $V_{OC}$ of $Ag_3TaS_4$, $Ag_3TaSe_4$, and $Ag_3TaTe_4$, respectively. The internal temperature increase under these V = $V_{OC}$ conditions is allocated primarily to non-radiative pathways, notably SRH recombination. This associated heating profile is maximum at the absorber surface far from the interface, with the values of $2.82 \times 10^9$ W/m$^3$ of $Ag_3TaS_4$, $8.78 \times 10^8$ W/m$^3$, and $2.11 \times 10^9$ W/m$^3$ of $Ag_3TaTe_4$. At V = $V_{OC}$ conditions, no net current flows and all photo-generated carriers recombine within the cell. The non-radiative heat generation rises with increasing distance from the window (CdS)/absorber ($Ag_3TaX_4$) interface, reflecting the progressively enhanced bulk SRH recombination. This trend arises from the higher concentration of structural defects in the absorber bulk, which promotes non-radiative carrier recombination and consequently elevates heat generation [54-55].

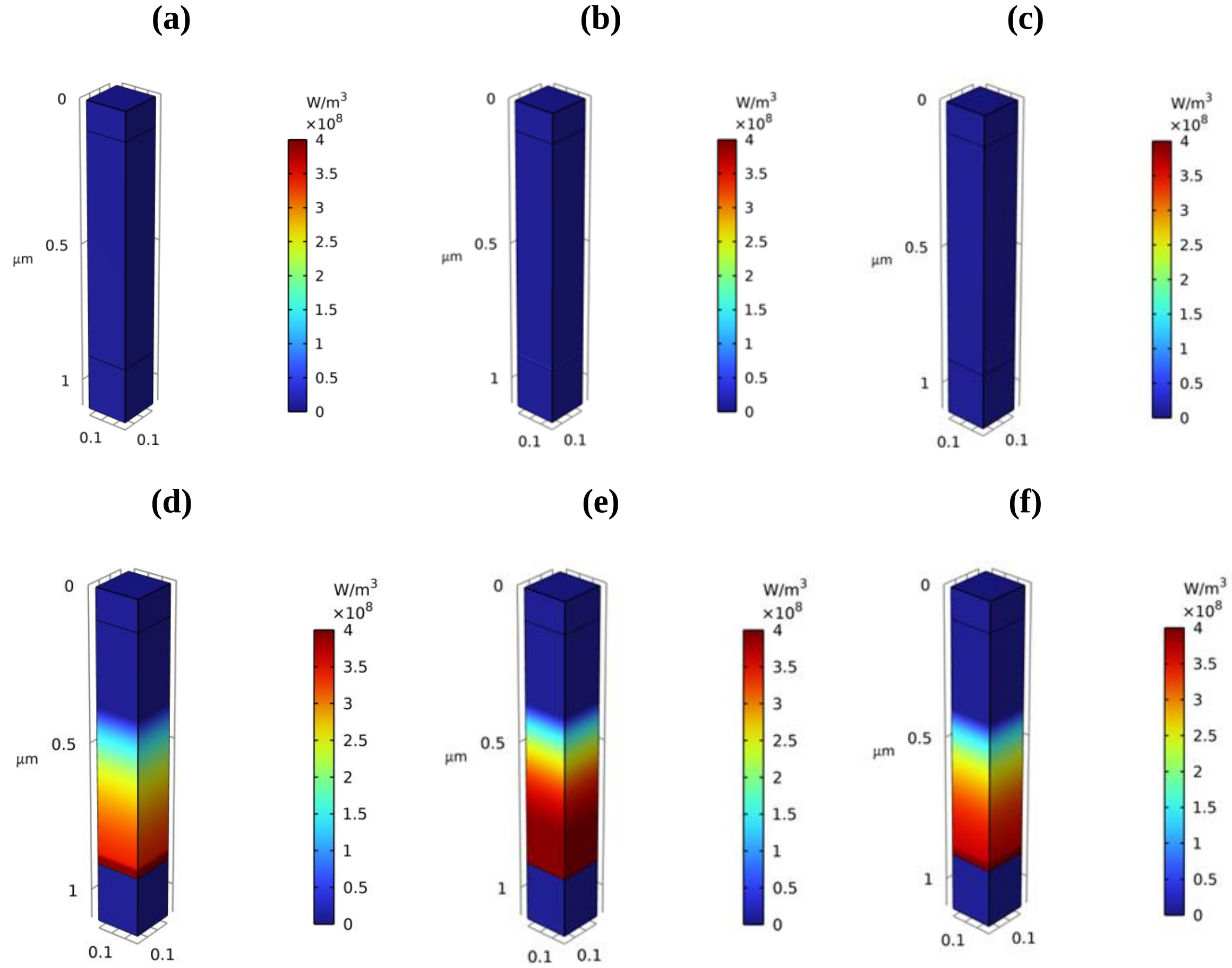

**Fig. 10:** Spatial arrangement of heat generation from recombination energy loss at V = 0 V of the (a) $Ag_3TaS_4$ (b) $Ag_3TaSe_4$ and (c) $Ag_3TaTe_4$ and at V = $V_{OC}$ of the (d) $Ag_3TaS_4$ (e) $Ag_3TaSe_4$ and (f) $Ag_3TaTe_4$-based solar cell devices.

### 3.7 Enhanced device functionality

Fig. 11(a) illustrates the performance enhancement in the $Ag_3TaX_4$-based SC with and without a $p^+$-GeS back surface field (BSF) layer.

In $Ag_3TaS_4$, incorporating a 0.2 µm thick back surface field (BSF) layer, doped at $10^{19}$ $cm^{-3}$, between the absorber and the back metal contact yields a substantial improvement in solar cell performance. Compared to the baseline device, which exhibits a $V_{OC}$ of 1.03, a $J_{SC}$ of 16.48 $mA/cm^2$, and an efficiency of 14.62%. The integration of the BSF layer elevates these metrics to 1.4V, 20.68 $mA/cm^2$, and 24.66%, respectively. Similarly, BSF also improves the all parameters for $Ag_3TaSe_4$ based solar cell from 0.98V to 1.19V, 21.87 $mA/cm^2$ to 27.00 $mA/cm^2$ and 18.40% to 28.1%, respectively. Without the GeS BSF layer, the voltage is 0.71V, $J_{SC}$ is 30.10 $mA/cm^{2,}$ and the efficiency is 17.85% for the $Ag_3TaTe_4$-based device. However, introducing a 0.2 µm thick and doped at $10^{19}$ $cm^{-3}$ $p^+$-GeS BSF layer between the absorber and back metal contact significantly enhances all key performance metrics. $V_{OC}$, $J_{SC}$, and PCE are increased to 0.88 V, 36.14 $mA/cm^2$, and 27.56%, individually, in $Ag_3TaTe_4$ [39]. This improvement can be attributed to an additional electric field formed at the $p$-$p^+$ interface between $Ag_3TaX_4$ and GeS layer, which repels minority carriers away from the back contact, thereby reducing recombination losses and increasing overall device performance [47].

Figure 11(b) demonstrates the improvement in quantum efficiency (QE) achieved by incorporating a GeS layer as a back-surface field (BSF). Prior to BSF integration, the QE begins to decline beyond approximately 500 nm, starting from values of 85% for $Ag_3TaS_4$, 86% for $Ag_3TaSe_4$ and 87% for $Ag_3TaTe_4$, with the cut-off wavelengths reaching 728 nm, and 826 nm, 1000 nm, respectively.

However, after incorporating the BSF layer, the QE is marginally enhanced, approaching 99% across all three materials. From the figure, it is visualized that QE is lower in the wavelength range of 400–500 nm for all materials. CdS window layer absorbs a large number of shorter wavelength photons. This absorption results in window gain, as a result, QE leads to lower. [56-57]. Beyond 520 nm wavelength, the QE peaks at over 99% as the absorber layer absorbs photons of all wavelengths of $Ag_3TaS_4$, $Ag_3TaSe_4$ and $Ag_3TaTe_4$. For $Ag_3TaS_4$, when wavelength exceed 728 nm, QE sharply decreases to zero. This value is 826 for $Ag_3TaSe_4$ and 1000 nm for $Ag_3TaTe_4$. This corresponds to the bandgap limit of each material, where photon energies fall below the absorption threshold [58].

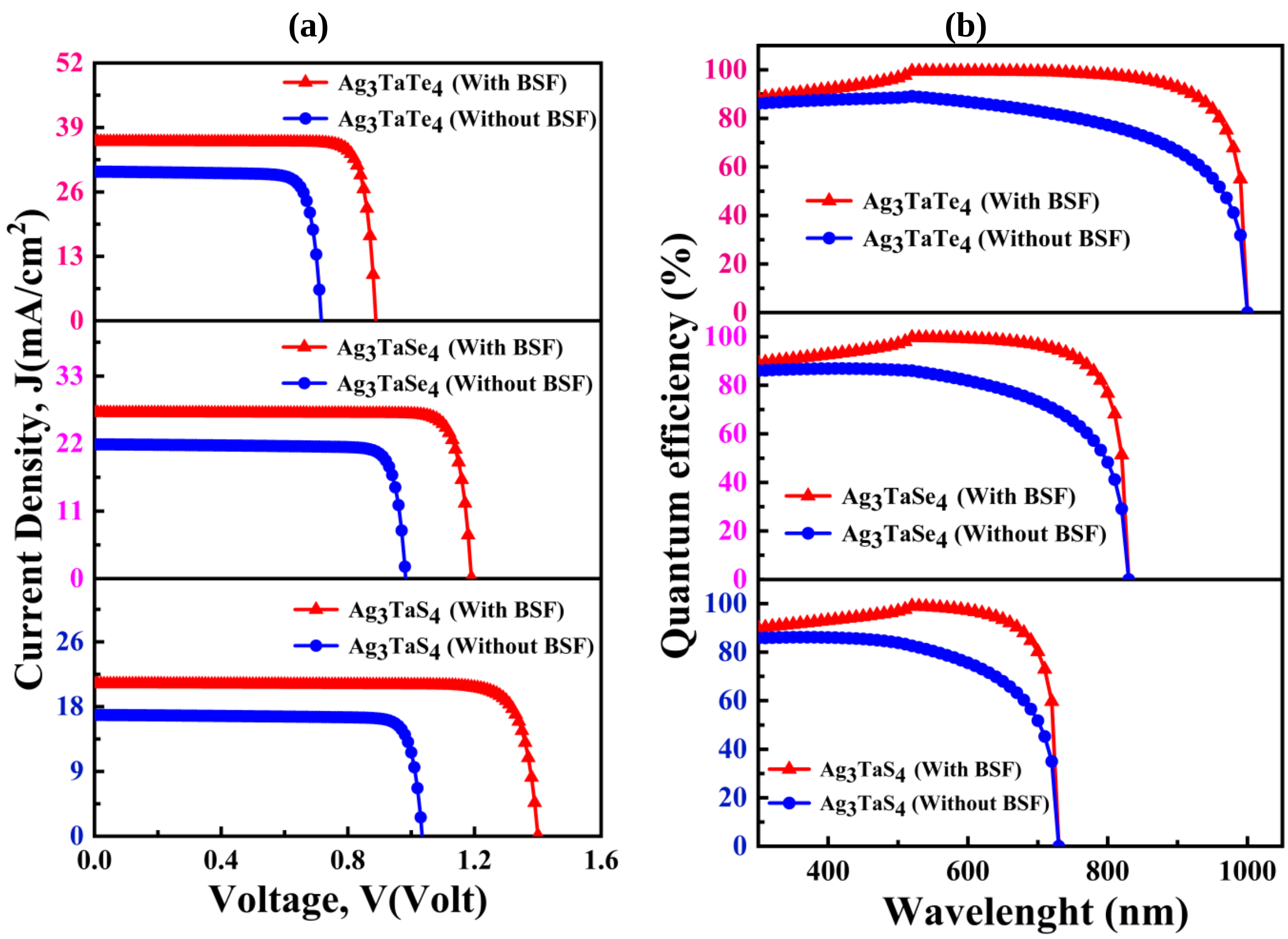

**Fig. 11:** (a) Voltage vs Current density characteristics and (b) Quantum efficiency vs wavelength characteristics of $Ag_3TaX_4$ (X = S, Se, Te)-based solar devices with and without BSF layer.

## 4. Conclusion

In this work, three dimensional n-CdS/p-$Ag_3TaX_4$ (X = S, Se, Te)/$p^+$-GeS heterojunction thin film solar cells have been designed and systematically investigated using the Semiconductor Module of COMSOL Multiphysics. The FEM simulations provided detailed insight into the carrier transport, electric-field distribution, recombination behavior, and thermal characteristics of the proposed devices, demonstrating the suitability of $Ag_3TaX_4$ ternary chalcogenides as promising absorber materials for next-generation photovoltaic applications. A comprehensive analysis is performed by varying the thickness, doping concentration, and defect density of the $Ag_3TaX_4$ (X = S, Se, Te) absorber, CdS window, and BSF layers. The results indicate the supremacy of the absorber layer properties on device performance, whereas moderate defect densities in the CdS and GeS layers produce only negligible changes in the PV characteristics. Under the optimized conditions, $Ag_3TaS_4$-, $Ag_3TaSe_4$-, and $Ag_3TaTe_4$-based SC devices exhibit the PCE of 24.66%, 28.1%, and 27.56%, respectively. The GeS BSF layer effectively suppresses rear-surface recombination by establishing a favorable p–$p^+$ electric field, thereby improving carrier collection and increasing the overall conversion efficiency by 10.04%, 9.70%, and 9.71%, for the $Ag_3TaS_4$-, $Ag_3TaSe_4$-, and $Ag_3TaTe_4$-based PV device, individually. The electrical and thermal analyses further revealed that low series resistance, high shunt resistance, and operation near room temperature are essential for maximizing device efficiency. Overall, this study about $Ag_3TaX_4$(X = S, Se, Te) compounds as highly promising absorber materials establishes and demonstrates the advantages for efficient thin film solar cells and of combining them with CdS and GeS in a 3D heterojunction architecture. The presented optimization strategy and thorough numerical investigation provide practical design guidelines for future experimental realization and further development of high-performance $Ag_3TaX_4$-based PV devices.

***Corresponding Authors:** E-mail: jak_apee@ru.ac.bd (Jaker Hossain).

**CRediT authorship contribution statement**

**Md. Nahid Hasan:** Writing – original draft, Methodology, Investigation, Formal analysis, Software, Data curation. **Tanvir Ahmed:** Writing – original draft, Visualization, Methodology, Validation, Formal analysis. **Md. Abdur Rashid:** Writing – original draft, Visualization, Methodology, Formal analysis. **Tanzina Rahman:** Writing – original draft, Visualization, Methodology, Formal analysis. **Dinesh Pathak:** Writing – original draft, Visualization, Methodology, Formal analysis. **Jaker Hossain:** Writing – review & editing, Writing – original draft, Visualization, Validation, Supervision, Methodology, Formal analysis, Conceptualization.

**Conflicts of Interest:** The authors have no conflicts of interest.

**Data availability:** Data will be available from the corresponding author upon reasonable request.

**Declaration of generative AI and AI-assisted technologies:** None of the authors use any AI or AI-assisted technologies in writing this manuscript.